\documentclass[12pt,draft]{article}

\usepackage{amsfonts} 
\usepackage{graphics}
\usepackage{amsmath}
\usepackage{amscd}
\usepackage{amssymb}
\usepackage{rlepsf}
\usepackage{color}

\usepackage{euscript}

\newtheorem{Theorem}{Theorem}[section]

\newtheorem{Definition}[Theorem]{Definition}

\newtheorem{Remark}[Theorem]{Remark}

\newtheorem{Lemma}[Theorem]{Lemma}

\newtheorem{Fundamental Theorem}{Fundamental Theorem}

\newenvironment{Proof}[1][Proof]{\textbf{#1.} }{\ \rule{0.5em}{0.5em}}

\def \C {\mathbb{C}}
\def \CH {{\mathrm{CH}}}
\def \d {\partial}
\def \e {\epsilon}

\def\l{\lambda}
\def \G {\Gamma}

\def \S {\Sigma}
\def \s {\scriptstyle}
\def \T {\mathrm{T}}
\def \TV  {{\mathrm{TV}}}
\def \W {\Omega}
\def \WRT {{\mathrm{WRT}}}

\begin{document}

\title{Invariants of Spin Networks Embedded in Three-Manifolds}

\author{ Jo\~{a}o  Faria Martins\footnote{Also at Departamento de Matem\'atica, Universidade Lus\'{o}fona de Humanidades e Tecnologia, Av do Campo Grande, 376, 1749-024, Lisboa, Portugal. }\\ \footnotesize\it  {Departamento de Matem\'{a}tica, Instituto Superior T\'{e}cnico,}\\{\footnotesize\it Av. Rovisco Pais, 1049-001 Lisboa, Portugal}\\ {\footnotesize\it jmartins@math.ist.utl.pt}\\ \\
Aleksandar Mikovi\'{c}\footnote{{Member of Grupo de F\'{\i}sica Matem\'{a}tica da Universidade de Lisboa, Av. Prof. Gama Pinto,
2, 1649-003 Lisboa, Portugal.
}} \\ 
\footnotesize \it{Departamento de Matem\'atica}\\ \footnotesize\it {Universidade Lus\'{o}fona de Humanidades e Tecnologia,}\\ \footnotesize \it{Av do Campo Grande, 376, 1749-024, Lisboa, Portugal}\\ \footnotesize \it {amikovic@ulusofona.pt}}

\date{\today}

\maketitle

\begin{abstract}
{We study the invariants of spin networks  embedded in a three-dimensional manifold which are based on the  path
  integral for $\mathrm{SU}(2)$ BF-Theory.} {These invariants  appear naturally in Loop Quantum Gravity, and have been defined as spin-foam state sums}. {By using the Chain-Mail  technique, we give a more general definition of these invariants, and show that the state-sum definition is a special case. This provides a rigorous proof that the state-sum invariants of spin
networks are topological invariants. } We
  derive various results about the BF-Theory spin network invariants, and we find a
  relation with the corresponding invariants defined from Chern-Simons Theory, i.e.
  the Witten-Reshetikhin-Turaev invariants. We also prove that the
  BF-Theory spin network invariants coincide with  V. Turaev's definition of
  invariants of coloured graphs embedded in 3-manifolds and thick
  surfaces, constructed by using  shadow-world evaluations. {Our framework
  therefore  provides a unified  view of these invariants}.
\end{abstract}

2000 Mathematics Subject Classification: 57M27; 57R56; 81T25.

\tableofcontents

\section{Introduction}
   {Spin networks, also called coloured graphs in this article, were invented by
R. Penrose \cite{P} in an attempt to find a combinatorial definition of
quantum geometry. These are trivalent graphs $\G$ whose edges
are coloured  with half-integers, i.e.  spins. Since each spin corresponds to an irreducible representation of  $\mathrm{SU}(2)$, one can consider graphs of arbitrary valence, by colouring  the vertices with appropriate intertwiners, see \cite{Ba}.  Each vertex of valence higher than three can be decomposed into  three-valent vertices; hence, without loss of generality,    {we} can restrict our attention to   three-valent spin networks.}

It was shown by C. Rovelli and L. Smolin that spin networks do  arise  
in a quantum theory of gravity (see \cite{RoS}), which is known as  ``Loop Quantum
Gravity'' (LQG). Each state $|\Psi\rangle$ in the LQG Hilbert space can be
expanded in the spin network basis $\{|\G\rangle\}$, which is in a one-to-one
correspondence with the set of  isotopy classes of closed spin networks $\G$ embedded in
a 3-manifold $M$\footnote{In LQG the 3-manifold $M$ represents the space, while the 4-manifold $M\times\bf R$ represents the spacetime.}. Explicitly  we have
$$ |\Psi\rangle = \sum_\G I_\Psi(M,\G)|\G\rangle\,, $$
where $I_\Psi(M,\G)$ is a certain invariant    {of   spin networks $\G$} embedded in $M$.

The invariant $I_\Psi$ can be represented in the Euclidean gravity case as a functional integral:
\begin{equation} \label{pi}I_\Psi(M,\G)= \int {\cal D} A \,W_{\G}[A] \Psi[A]\,,\end{equation}
called the loop transform of $\Psi[A]$, where $A$ is a real $\mathrm{SU}(2)$ connection on $M$, $W_{\G}[A]$ is the spin network wave-functional  (the product of the holonomies along the edges of $\G$ contracted    {by  using} the intertwiners assigned to the vertices of $\G$, see \cite{Ba}) and $\Psi[A]$ is a diffeomorphism and gauge invariant functional of $A$ which satisfies the Hamiltonian constraint equation. In the Minkowski gravity case,
the connection $A$ is complex and the loop transform has a more complicated path-integral representation, see \cite{Ko2}.

Kodama \cite{Ko} has shown that an exponent of the Chern-Simons functional
\begin{equation} \Psi_{\mathrm{CS}} [A] = \exp\frac{1}{\lambda}\int_M \mathrm{Tr}\left(A\wedge dA + \frac{2}{3}A\wedge A\wedge A \right)\label{kod}\end{equation}
satisfies the Hamiltonian constraint equation, where $\lambda$ is the cosmological constant. The formula (\ref{pi}) then implies, see \cite{MS,M1,FS}, that  $I_\l (M,\G)\doteq I_{\Psi_{\mathrm{CS}}}(M,\G)$ can be related by an analytic continuation    {$\l\mapsto i\l$} to the Witten-Reshetikhin-Turaev Invariant $Z_\WRT(M,\G)$ of the pair $(M,\G)$. Furthermore, it was conjectured that $\lambda\propto k^{-1}$ where $k/2$ is the maximal spin used in the construction of  the WRT Invariant; see \cite{S,MS}. 

When $\lambda =0$ then it is possible to show that the    {functional:}
$$ \Psi_0 [A] =\delta[F]= \int {\cal D}B \, e^{ i\int_M Tr\left(B\wedge F \right)} \,,$$
 satisfies the Hamiltonian constraint equation,
where $F=dA + A\wedge A$ and $B$ is a one-form,    {see \cite{M1}.} Therefore the path integral: 
$$ I_0(\G,M)= \int {\cal D} A \,{\cal D}B \,W_{\G}[A] \, e^{ i\int_M Tr\left(B\wedge F \right)}$$
provides us with a physical definition  of quantum invariants  of 3-dimensional oriented manifolds based on $\mathrm{SU}(2)$    {BF-Theory.} The path integral $I_0$ also provides a set of observables for the theory of flat $\mathrm{SU}(2)$ connections on $M$.    { This theory can  be also considered as  a generalisation of  three-dimensional Euclidean General Relativity.}

The invariant $I_0$ can be defined by using the spin foam techniques
\cite{M1,M2}, also see \cite{O}. Let $\Delta (M)$ be a triangulation of the
oriented 3-manifold $M$. Let $\Delta^*(M)$ be the dual cell complex; see
\cite{RS}. Let $H_-$ be the 3-dimensional index 1 handlebody obtained by
thickening the 1-skeleton of $\Delta^*(M)$. The spin network $\G$  can be
inserted into $H_-$    {in a way} such that each 0-handle of $H_-$ contains
at most one vertex of $\G$,     {while  each edge of $\G$}    {goes along
  some  1-handle of $H_-$, parallel to its core}. The invariant $I_0$ can be then expressed as
\begin{equation} 
 I_0(M,\G)= \int \prod_{l\in L} dg_{l} \,W_{\G}(g)\prod_{f\in F} \delta
 (\hat{g}_f)\, ,\label{pid}\end{equation}
where $L$ is the set of edges of $\Delta^*(M)$, $F$ is the set of faces of $\Delta^*(M)$, $g_l$'s are the dual edge holonomies and  $\hat{g}_f =\prod_{l\in\partial f} g_l$. The spin network wavefunction $W_\G (g)$ is given by:
$$ W_\G (g) = \Big{\langle}\bigotimes_{l\in
  L(\G)}D^{(j_l)}(g_l)\,\Big{|}\,\bigotimes_{v\in V(\G)}\iota_v
\Big{\rangle}\, ,$$
   {where $L(\G)$ is the subset of  edges of the dual complex $\Delta^*(M)$
   which  pass through the same 1-handles of $H_-$ as the edges of $ \G$,  $j_l$ is the spin assigned to the corresponding  edge of $\G$,}  $D^{(j_l)}(g_l)$ is the {appropriate} $\mathrm{SU}(2)$ representation matrix, $V(\G)$    {is the subset of vertices of $\Delta^*(M)$ which    {are contained in the same} 0-handles of $H_-$ as the vertices of $\G$}, $\iota_v$  are the corresponding intertwiners and $\langle|\rangle$ denotes the  contraction of appropriate indices.

Performing    {the} group integrations in (\ref{pid}) is equivalent to using the skein
relations for the $\mathrm{SU}(2)$ group; see \cite{Ba,O,M2}. Hence one
obtains a state sum in terms of classical $nj$ symbols where $n\ge 6$
\cite{M1,M2},    {which reduces to the  Ponzano-Regge Model \cite{PR} when the graph $\G$ is empty}.    {These infinite sums can be regularised} by passing to the quantum $\mathrm{SU}(2)$ group at a root of unity, which then gives a Turaev-Viro type state sums containing  quantum $nj$ symbols \cite{M1}. We will
denote this invariant by $Z_\CH(M,\G)$, or $Z_\CH(M,\G;j_1,...,j_n)$ if we
want to emphasise the colourings of the edges of $\G$. 

In this paper we are going to show that the state sum $Z_\CH(M,\G)$ can be obtained    {in the context of}  J. Roberts' Chain-Mail formalism \cite{R1,R2}, which then provides a proof that $Z_\CH(M,\G)$ is a well-defined invariant of    { coloured framed graphs $\G$} embedded in an oriented    {closed 3-manifold $M$.} The Chain-Mail approach also provides a simple and natural definition of    {these} spin network invariants, as well as an easier way to calculate    {them.}

We will also prove that
\begin{equation} Z_\CH(M,\G) = Z_\WRT(M,\G)Z_\WRT(\overline{M}),\label{bfcs}\end{equation}
where $\overline{M}$ denotes  the oriented 3-manifold $M$ with the opposite orientation. Hence the equation (\ref{bfcs}) gives a relation between     {BF-Theory} spin network invariants and the corresponding    {Chern-Simons Theory} invariants.

The relation (\ref{bfcs}) also implies that the invariant $Z_\CH(M,\G)$ coincides (apart from normalisation factors) with  the graph invariants    {$Z_\TV(M,\G)$} defined in a combinatorial way by V. Turaev in \cite{T3,T4}; see also \cite{BD,KS}. This is a surprising result, given the completely different ways of constructing these invariants.    {The definition} of Turaev's Invariant    {$Z_\TV(M,\G)$} is quite complicated. One has to
excise from $M$ a regular neighbourhood $n(\G)$ of  $\G$,  and what is left of $M$
has to be triangulated. Then a Turaev-Viro type state sum is constructed, having  some  extra terms obtained  by natural shadow-world evaluations  at the boundary of $n(\G)$. The construction of    {$Z_\TV(M,\G)$} will be outlined in this article, and we will also provide a direct proof that    {$Z_\TV(M,\G)$} coincides with $Z_\CH(M,\G)$, apart from normalising factors. 

The invariant $Z_{\CH}(M,\G)$ is also well defined for manifolds with
boundary, a fact which we will prove carefully;   {compare with \cite{BP}.}  This makes it possible to define $Z_\CH(M,\G)$ for the case of thick surfaces $M=\S\times I$. In this case it is possible obtain a combinatorial definition of  $Z_\CH(M,\G)$, making use of natural handle decomposition of $\S \times I$,  yielding  exactly  the shadow world definition in \cite{T3,T4}. 

This framework also provides an alternative definition of the Turaev-Viro
Invariant $Z_\TV(M)$ for compact manifolds with boundary, see \cite{KMS}. In
fact, what we prove is that the equivalent    {J. Roberts'} Chain-Mail
Invariant $Z_\CH(M)$ extends directly to manifolds with boundary;  {cf. \cite{BP}}.

This article should be compared with \cite{BGM} where a different set of invariants of graphs embedded in oriented 3-manifolds is defined. In fact the techniques of proof in both articles are very similar, however adapted to different invariants. We will refer frequently to results obtained in \cite{BGM} and \cite{R1,R2} in the course of this article.  See also \cite{B,GI,B2,B3}.

\section{Preliminaries}
\subsection{Chain-Mail}
We use the normalisations of \cite{BGM}. The Turaev-Viro Invariant $Z_\TV$ of 3-dimensional closed manifolds can be defined in the fashion shown in this section, due to    { J. Roberts;}  see \cite{R1,R2}.
 For an introduction to handle decomposition of manifolds, we refer the reader to \cite{RS}, \cite{GS} and \cite{K}.  

One advantage of    { Roberts'} approach is that it makes the introduction of observables very natural, and all proofs can be obtained in a geometrical, rather than combinatorial way. The extension of the results to manifolds with boundary is immediate    {within} this framework. 

\subsubsection{Generalised Heegaard Diagrams}
 Let $M$ be a closed oriented (piecewise-linear) 3-manifold. Choose a handle decomposition of $M$. Let $H_-$ be the union of the 0- and 1-handles of $M$. Let also  $H_+$ be the union of the 2- and 3-handles of $M$. Both $H_-$ and $H_+$ have natural orientations induced by the orientation of $M$. There exist two non-intersecting naturally defined  framed links $m$ and $\e$ in $H_-$. The second one is given by the attaching regions of the 2-handles of $M$ in $\d H_-=\d H_+$, pushed inside $H_-$, slightly. On the other hand, $m$  is given by the meridional disks of the 1-handles, in other words by the belt-spheres of the 1-handles of $M$, living in $\d H_-$.  The sets of curves $m$ and $\e$  in $H_-$ have natural framings, parallel to $\d H_-$. The triple $(H_-,m,\e)$ will be called a generalised Heegaard diagram of the oriented closed 3-manifold  $M$. 

Generalised Heegaard diagrams are also defined for compact 3-manifolds with a non-empty boundary. The only difference is that $\d H_+$ is not equal to $\d H_-$ anymore.

\subsubsection{Skein Theory and the $\W$-element}
Let $L$ be a framed link in $S^3$. Consider an integer parameter $r \ge 3$, and let $A=e^{\frac{i \pi} {2r}}$ and $q=A^2$. Let also    { $\dim_q j=\frac{q^{2j+1} - q^{-2j-1}}{q-q^{-1}}$.}  

If the components of the framed link  $L$ are assigned spins $j_1,j_2,...,j_n
\in \{0,1/2,...,(r-2)/2\}$, then we can consider the value $\left
  <L;j_1,...,j_n\right> \in \C$    {(the skein space of $S^3$),} obtained by evaluating the coloured Jones polynomial at $q$; see \cite{KL}. For example, if    {$\bigcirc_0$} is the unlink with 0-framing, then    {$\left <\bigcirc_0;j\right >=\dim_q j$. Colourings by spins should not be confused with framing coefficients when drawing coloured links. It should be obvious from the context which colouring is meant in each case.}

\begin{figure}
\centerline{\relabelbox 
\epsfysize 2.5cm
\epsfbox{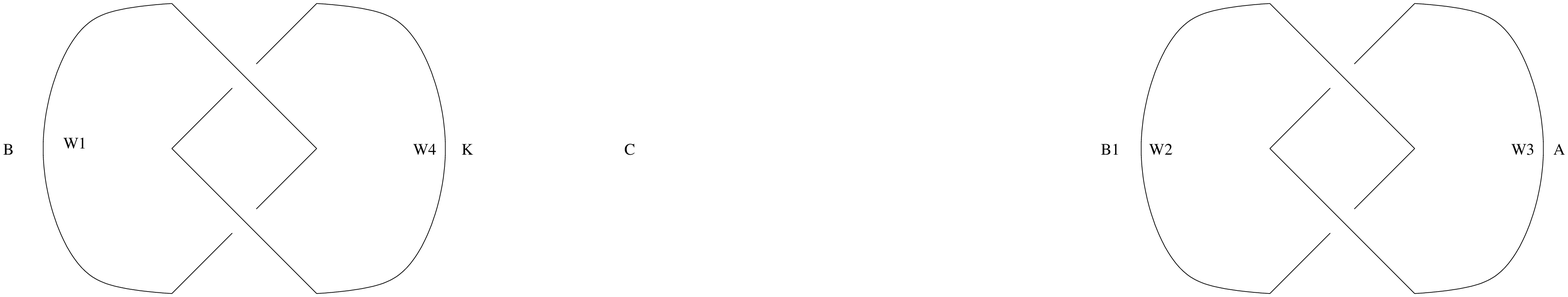}
\relabel{W1}{$\s{\W}$}
\relabel{W2}{$\s{a}$}
\relabel{W3}{$\s{b}$}
\relabel{W4}{$\s{\W}$}
\relabel{B}{$\Big{\langle}$}
\relabel{B1}{$\Big{\langle}$}
\relabel{K}{$\Big{\rangle}$}
\relabel{A}{$\Big{\rangle}$}
\relabel{C}{$\s{=\displaystyle{\sum_{a,b}\dim_q(a)\dim_q(b)}}$}
\endrelabelbox } 
\caption{\label{ex} Evaluation of the Hopf Link with both strands coloured with $\W$.    {All strands are given the blackboard framing.}}
\end{figure}

\begin{figure}
\centerline{\relabelbox 
\epsfysize 2cm
\epsfbox{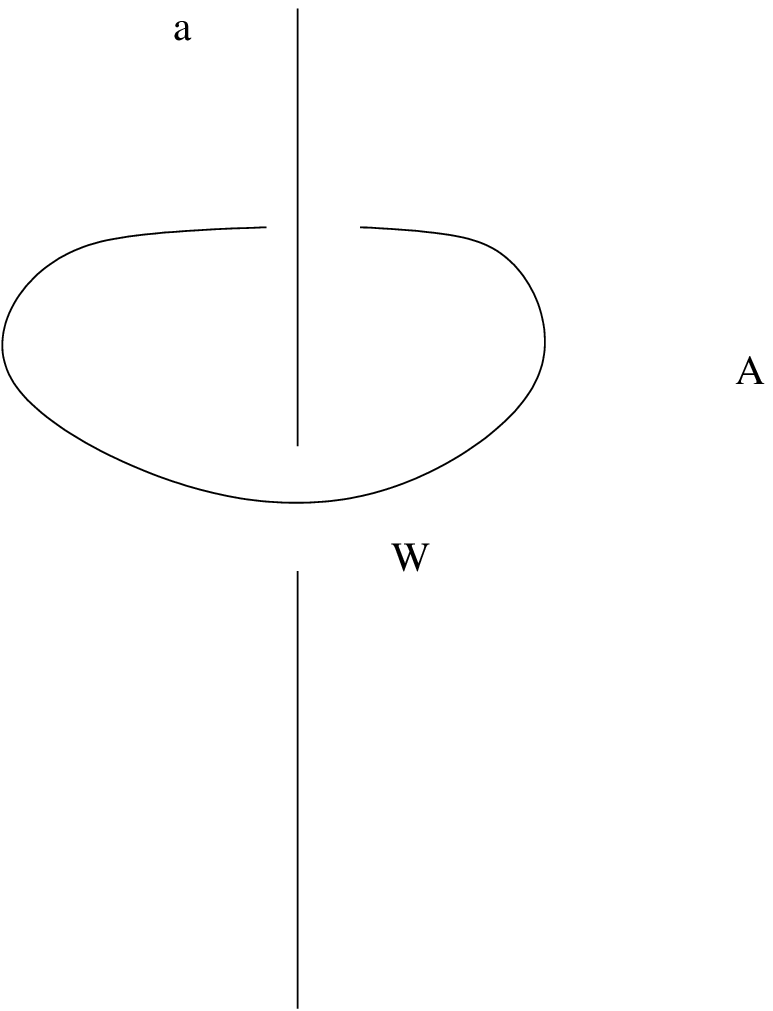}
\relabel{W}{$\s{\W}$}
\relabel{a}{$\s{a}$}
\relabel{A}{$=N\delta(a,0)$}
\endrelabelbox }\caption{\label{kill} Lickorish Encircling Lemma for the case of one strand: the ``Killing an $\W$'' property. All networks are given the blackboard framing. }
\end{figure}

We can in general consider the case in which the components of $L$ are assigned linear combination of spins, {with multilinear dependence on the colourings of each strand of $L$}. The most important of these linear combinations  is the ``$\W$-element'' given by: 
$$   {\W= \sum_{j=0}^{\frac{r-2}{2}}   \dim_q(j) j.}$$ 
   {It is understood that:}
$$   {\left<\bigcirc; \W  \right>=\sum_{j=0}^{\frac{r-2}{2}}   \dim_q(j)  \left< \bigcirc ;j \right>,}$$
   {where ``$\bigcirc$'' represents some link component.}
For example for the evaluation of the Hopf Link with both strands coloured by the $\W$-element, one would proceed as in figure \ref{ex}. By the ``Killing an $\W$'' property (see figure \ref{kill}) it actually follows that the end result is the element $N\in \mathbb{R}$, defined below.   

A sample of the beautiful properties satisfied by  the $\W$-element appears in \cite{R1,R2,L,KL}. See figures \ref{kill}, \ref{Lic2}, \ref{lic}  and \ref{lic4} for different consequences of the important Lickorish Encircling Lemma satisfied by a 0-framed unknot coloured by the $\W$-element; see \cite{L}. Another    {extremely} important property is the    {handleslide invariance} property, shown in the references above.

Define    {$\left <\bigcirc_0;\W\right >=N$} and     {$\left <\bigcirc_1;\W\right >=\kappa\sqrt{N}$}, the evaluation of the 0- and 1-framed unknots coloured with the $\W$-element. Therefore we have that    { $N=\sum_{j=0}^{(r-2)/2} (\dim_q j)^2=\frac{2}{r} \sin \left ( \frac \pi r \right)$. }On the other hand     {$\kappa= A^{-3-r^2}e^{-\frac{i \pi}{4}}$}, and      {$\left < \bigcirc_{-1}\right >=\sqrt{N}\kappa^{-1}$;} see \cite{R1,R2}. 

Note that the evaluation $\left <L;j_1,...,j_n\right>$ can be extended to 3-valent framed graphs $\G$ whose edges are coloured with spins, {or linear combinations of them}. We use the normalisations of \cite{KL} for the 3-valent vertex. By a framed graph we mean a finite graph whose edges and vertices are extended to small 2-disks and bands. These are called fat graphs in \cite{T3}. We will only consider framed graphs for which their associated surface with boundary is orientable. 

   {If $\G$ and $\G'$ are non-intersecting framed graphs  in $S^3$, coloured
  with the spins $j_1,...,j_n$ and $i_1,...,i_m$, then there exists a
  colouring of their disjoint union $\G \cup \G' \subset S^3$, and its 
  evaluation in the  skein space
  of $S^3$ is denoted by $\left < \Gamma \cup \Gamma ';j_1,...,j_n, i_1,...,i_m \right>$. We will use
  this type of  notation frequently in this article.}
\subsubsection{The Chain-Mail Invariant}
Let again $M$ be a {connected} 3-dimensional closed oriented piecewise linear  manifold. Consider a generalised Heegaard diagram $(H_-,m,\e)$, associated with a handle decomposition of $M$. Give $H_-$ the orientation induced by the orientation of $M$.  Let $\Phi\colon    H_- \to S^3$ be an orientation preserving embedding. Then the image of the links $m$ and $\e$ under $\Phi$ defines a link $\CH(H_-,m,\e,\Phi)$ in $S^3$, called the ``Chain-Mail Link''. 
   {J. Roberts} proved  that the evaluation 
$\left < \CH(H_-,m,\e,\Phi); \W \right>$
of the Chain-Mail Link coloured with the $\W$-element  is actually independent of the orientation preserving embedding $\Phi\colon   H_- \to S^3$; see \cite{R1}, proposition $3.3$. 

The Chain-Mail Invariant of $M$    {(due to    {J. Roberts})} is defined  as:
$$Z_\CH(M)=N^{g-K-1}\left < \CH(H_-,m,\e,\Phi);\W \right>,$$
where $K$ is the number of components of the Chain-Mail Link and $g$ is the genus of the Heegaard surface  $\d H_-$. Therefore $K=n_1+n_2$ and $g=n_1-n_0+1$, where $n_i$ is the number of $i$-handles of $M$.  It is proved in \cite{R1,R2} that this    {Chain-Mail Invariant}  coincides with the Turaev-Viro Invariant $Z_\TV(M)$ of $M$; see \cite{TV}. We will go back to this later; see \ref{TViro}.

Even though we need an orientation of $M$ to define $Z_\CH$, it is possible to prove that the Chain-Mail Invariant is orientation independent. In fact, the equivalent Turaev-Viro Invariant  is  defined also for non-orientable closed  3-manifolds. 

\subsection{The Witten-Reshetikhin-Turaev Invariant}
The main references now are \cite{L} and \cite{RT}.
Let $M$ be an oriented {connected} closed $3$-manifold. Then $M$ can be
presented by surgery on some framed link $L \subset S^3$, up to orientation
preserving diffeomorphism. Any    {framed} graph $\G$ in $M$ can be pushed away from the areas where the surgery is performed, and therefore any pair $(M,\G)$, where $\G$ is a trivalent framed graph in the oriented closed  3-manifold $M$ can be presented as a pair $(L,\hat{\G})$, where $\hat{\G}$ is a framed trivalent graph in $S^3$, not intersecting $L$.    

The Witten-Reshetikhin-Turaev Invariant of a pair $(M,\G)$, where the framed graph $\G$  is coloured with the spins $j_1,...,j_n$,  is defined as:
$$Z_\WRT(M, \G;j_1,...,j_n)=N^{-\frac{m+1}{2}}\kappa ^{-\sigma(L)}\left <L \cup \hat{\G};\W,j_1,...,j_n \right>.$$
Here $\sigma(L)$ is the signature of the linking matrix of the framed link $L$.
This is an invariant of the pair $(M,\Gamma)$, up to orientation preserving diffeomorphism.
Note that, in contrast with the Turaev-Viro Invariant, the  Witten-Reshetikhin-Turaev Invariant  is sensitive to the orientation of $M$. {If $M$ is an oriented 3-manifold, we represent the manifold with the reverse orientation by $\overline M$. }

Some immediate  properties of the  Witten-Reshetikhin-Turaev Invariant
are the following:
\begin{Remark}\label{WRTp}
We have:

$$ Z_{\WRT}(S^3)=N^{-1/2}, \qquad Z_{\WRT}(S^2\times S^1)=1,$$
$$ Z_{\WRT}(\overline M,\Gamma)=\overline{Z_{\WRT}(M,\Gamma)},$$
$$    {Z_{\WRT}\big((P,\Gamma)\# (Q,\Gamma')\big )= Z_{\WRT}(P,\Gamma)Z_{\WRT}(Q,\Gamma')N^{\frac{1}{2}}.}$$
\end{Remark}
   {Here $M$, $P$ and $Q$ are oriented closed    {connected} 3-manifolds. In addition,   $\G$
  and $\G'$ are coloured graphs embedded in $P$ and $Q$.  It is understood
  that the connected sum $P \# Q$ is performed away from $\G$ and $\G'$. It is
  easy to see that this uniquely defines a    {framed} graph in    {$P \#
  Q$}, up to orientation preserving diffeomorphism.  To prove the third
  equation we need to use}  the fact that    {$N$ is a  real number.}

The invariant $Z_\WRT$ can be extended to manifolds with an arbitrary number of components by doing the product of the invariant $Z_\WRT$  of  each component; see \cite{BGM}.

Given oriented closed    {connected} 3-manifolds    {$P$ and $Q$}, we define    {$P\#_n Q$} in the following way; see \cite{BGM}. Remove $n$ 3-balls from    {$P$ and $Q$}, and glue the resulting manifolds    {$P'$ and $Q'$} along their boundary in the obvious way,    {so that the final result is an oriented manifold.}    {We denote it by:} $$   {P\#_n Q= P' \bigcup_{\d P' = \d Q'} Q'.}$$ 
   {This definition extends to the non-closed case in the obvious way.}
   {It is easy to see that}    {$P \#_1 Q=P \# Q$,} and     {$P \#_n Q$} is diffeomorphic to    {$(P \# Q)\# (S^1 \times S^2)^{\# (n-1)}$, if $n>1$.}
  
From remark \ref{WRTp} it follows immediately that: 
\begin{multline}\label{refer} 
Z_\WRT(P \#_n Q, \Gamma \cup \Gamma';j_1,...,j_p,i_1,...,i_m)\\=   
Z_\WRT(P, \Gamma;j_1,...,j_p)Z_\WRT( Q,  \Gamma';i_1,...,i_m)N^{\frac{n}{2}}.
\end{multline}
 Here $P$ and $Q$ are closed oriented 3-manifolds. In addition, $\G$ and $\G'$
 are graphs in $P$ and $Q$, coloured with the spins $j_1,...,j_p$ and
 $i_1,...,i_m$, respectively. As before, it is implicit that the multiple
 connected sum $P \#_n Q$ is performed away from $\G$ and $\G'$. This defines
 a graph $\Gamma \cup \Gamma'$ in $P \#_n Q$ up to orientation preserving
 diffeomorphism.    { This basically follows from the Disk Theorem (see \cite[3.34]{RS}), and the fact that graph complements in connected 3-manifolds are connected.}

\section{Spin Network Invariants}

\subsection{Definition of the Invariants}

Let $M$ be a closed connected oriented piecewise linear 3-manifold. Let also $\Gamma$ be a framed trivalent graph embedded in $M$, whose edges are coloured with  the spins $j_1,...,j_n \in \{0,1/2,...,(r-2)/2\}$.  There exists a handle decomposition of $M$ such that $\G$ is contained in $H_-$,  the union of the 0- and 1-handles of $M$. Let $(H_-,m,\e)$ be the associated generalised Heegaard diagram of $M$. Note that $H_-$ is connected.  Choose an orientation preserving embedding $\Phi\colon    H_- \to S^3$. The framed graph $\G$ can be added to  the Chain Mail Link $\CH(H_-,m,\e,\Phi)$, by using the map $\Phi$.

\begin{Definition}
The Chain-Mail Expectation    {Value} is by definition:
$$Z_{\CH}(M,\Gamma;j_1,...,j_n)=N^{g-K-1}\left<\CH(H_-,m,\e,\Phi) \cup \Phi(\G);\W,j_1,...,j_n \right>,$$
where,    {as before}, $K$ is the number of components of    {the Chain-Mail Link} $\CH(H_-,m,\e,\Phi)$ and $g$ is the genus of the Heegaard surface $\d H_-$.    {Recall} that $g=n_1-n_0+1$ and $K=n_1+n_2$, where $n_i$ is the number of $i$-handles of $M$.
\end{Definition}
The value of $Z_{\CH}(M,\Gamma;j_1,...,j_n)$ actually does not depend on the orientation preserving embedding $\Phi\colon    H_- \to S^3$. This follows from the same argument  of the proof of proposition $3.3$ of \cite{R1}. In fact, $Z_{\CH}(M,\Gamma;j_1,...,j_n)$ stays the same after births or deaths of $0-1$ pairs of handles, now by the same argument of the proof of theorem 3.4 of \cite{R1}. In particular, we can alter the handle decomposition of $M$ to a handle decomposition of $M$ with a unique $0$-handle, without changing $Z_{\CH}(M,\Gamma;j_1,...,j_n)$. This fact will be of    {prime} importance.

 The independence  of the Chain-Mail Expectation Value on a general handle decomposition of $M$ containing $\G$ in the union of $0$- and $1$-handles of $M$ can be proved as Theorem 1 of \cite{BGM}.

 {We can suppose  that $M$ has a unique $0$-handle.  In  this case,  the signature  $\sigma( \CH(H,m,\e,\Phi))$ of the linking matrix of the Chain-Mail Link  is zero {(see \cite[proof of theorem 3.7]{R1}, \cite[4.2]{BGM} or \ref{boundary})}. Therefore it follows that $Z_{\CH}(M,\Gamma,j_1,...,j_n)$ is the Witten-Reshetikhin-Turaev Invariant of the pair graph-manifold defined, by using surgery,  from  $(\CH(H_-,m,\e,\Phi),\Phi(\G))$, times
$N^{-n_0-n_2+\frac{n_1+n_2+1}{2}}=N^{-n_3/2}$, where $n_i$ is the number of
$i$-handles of $M$; see    {\cite[proof of theorem 1]{BGM}} and \ref{boundary}. Note that  the Euler characteristic of any closed orientable 3-manifold is zero, thus $n_0-n_1+n_2-n_3=0$. }

On the other hand (recall that  we  suppose that the chosen handle decomposition of $M$ has a unique 0-handle), by the same argument appearing in the proof of theorem 1 of \cite{BGM}, the pair $(\CH(H_-,m,\e,\Phi),\Phi(\G))$ is a surgery presentation of  the graph $\G$ inside the manifold $M \#_{n_3} \overline{M}$, considering the inclusion of $\G$ into $M$ (not $\overline{M}$);    {see lemma \ref{Sur} for an alternative proof of this important fact.}

Therefore: 
$$Z_{\CH}(M,\Gamma;j_1,...,j_n)=N^{-n_3/2} Z_{\WRT} (M \#_{n_3} \overline{M}, \G;j_1,...,j_n),$$
thus, by equation (\ref{refer}), it follows that:
$$Z_{\CH}(M,\Gamma;j_1,...,j_n)=     Z_{\WRT} (M, \G;j_1,...,j_n)   {Z_{\WRT} (\overline{M})}.$$
In particular:
\begin{Theorem}
Let $M$ be a closed  {connected} oriented piecewise linear manifold {of
  dimension 3}. Let also $\G$ be a framed trivalent graph embedded in $M$,
with its edges coloured with the spins $j_1,..,j_n \in
\{0,1/2,...,(r-2)/2\}$. The Chain-Mail Expectation Value
$Z_{\CH}(M,\G;j_1,...,j_n)$    {depends on neither   the handle decomposition of $M$ containing $\G$ in the union of $0$- and $1$-handles, nor on the orientation preserving embedding $\Phi\colon    H_- \to  S^3$,} and    {it} is therefore an invariant of the pair $(M,\G)$, up to orientation preserving diffeomorphism. In fact:
$$Z_{\CH}(M,\Gamma;j_1,...,j_n)=     Z_{\WRT} (M, \G;j_1,...,j_n)   \overline{Z_{\WRT} ({M})}.$$
\end{Theorem}
This theorem is ultimately due to V. Turaev, though it did not appear exactly in this form. See \cite{T1} and \cite[Theorem 7.3.1]{T4}.  We will go back to this issue  later; see \ref{triang}.

   {The construction given in \cite{M1,M2} corresponds to the  special case  when each 0-handle of $M$ contains at most one vertex of $\G$, while each edge of $\G$ goes along some 1-handle of $M$. Notice that this restriction does not appear in this article.}

Note that the {spin network invariant} $Z_{\CH}(M,\Gamma;j_1,...,j_n)$ {is} orientation dependent. In fact $Z_{\CH}(\overline{M},\Gamma;j_1,...,j_n)=\overline{Z_{\CH}(M,\Gamma;j_1,...,j_n)}$; see remark \ref{WRTp}. This is in sharp contrast with the observables defined in \cite{B,BGM}.

\subsubsection{Manifolds with Boundary}\label{boundary}

Let $M$ be a compact {connected} oriented piecewise-linear 3-manifold, with a non-empty boundary. Consider a
handle decomposition of $M$. As before, let $H_-$ be the union of the
0- and 1-handles of $M$. Let $\G$ be a {framed} graph embedded in $M$, such
that $\G \subset H_-$. Suppose that the edges of $\G$ are coloured
with the spins $j_1,...,j_n$.

 Consider an orientation preserving embedding $\Phi\colon    H_- \to S^3$. We
   define the Chain-Mail Expectation Value in an analogous way to the closed case:
$$Z_{\CH}(M,\Gamma;j_1,...,j_n)=N^{g-K-1}\left<\CH(H_-,m,\e,\Phi) \cup \Phi(\G);\W,j_1,...,j_n \right>.$$
This is independent of the orientation preserving    {embedding} $\Phi\colon   H_-
\to S^3$, for the same reason why it is in the closed case. Recall that $g$ is the genus of $\d(H_-)$, and $K$ is the number
of components of the Chain-Mail Link.  

Note that by the same argument as in the case of closed manifolds,  we
can suppose (and we will) that $M$ has a handle decomposition with a unique
0-handle. 
\begin{Lemma}\label{Sur}
Suppose that  $M$ has a unique 0-handle. Then it follows that  the pair $(\CH(H_-,m,\e,\Phi) ,
\Phi(\G))$ is a surgery presentation of the graph $\Gamma$ inside the
boundary of the    {compact} 4-manifold $M^{(2)} \times I$,    {where $I=[0,1]$.} Here $M^{(2)}$
is the union of the 0-,1- and 2-handles of $M$, and   $\G$ is embedded    {as a copy $\G \times \{0\}$ } in $M^{(2)} \times    {\{0\}} \subset \d (M^{(2)} \times I)$, in the obvious way. 
\end{Lemma}
\begin{Proof} {\bf (sketch)}
Turn the meridian curves $m$ of the $1$-handles of  $H_-$ into dotted circles, as in \cite{K};
see also \cite{GS}. Similarly to   \cite[proof of theorem 3.7]{R1},
consider the 4-manifold $W$ obtained from $D^4$ by looking at  the
Chain-Mail Link  $\CH(H_-,m,\e,\Phi)\subset S^3$  as being a Kirby diagram. In other
words, we attach 4-dimensional 1-handles to the dotted circles of the
Chain-Mail Link (after turning each of them into a disjoint union of 3-disks, {paying attention to the remaining components of the Chain-Mail Link}), and 2-handles to the remaining curves.  Then $W$ is
diffeomorphic to $M^{(2)} \times I$, where $M^{(2)}$ is the union of
the 0-, 1- and 2-handles of $M$. See for example \cite[proof of
theorem 3.7]{R1} or \cite[4.6.8]{GS}. Under the natural diffeomorphism
$W \to M^{(2)} \times I$, the region in $S^3\subset D^4$ inside the
unique 0-handle of $\Phi(H_-) \subset S^3$ goes exactly to the
corresponding area in    {$H_- \subset M^{(2)} \times \{0\}\subset \d (M^{(2)} \times I)$}; see \cite[4.6.8]{GS} and \cite{BP}.  This finishes the proof.   
\end{Proof}

{Now, we will use the fact  (see \cite{R1,R2}) that the  signature of the linking matrix of the Chain-Mail Link $\CH(H_-,m,\e,\Phi)$  is zero. This follows from the  fact that the Chain-Mail Link  defines a Kirby diagram for $M^{(2)}\times I$    {(a manifold of zero signature)} and lemma 3 of \cite{BGM}; a result included in the discussion in \cite{R1,R2}. This permits us to conclude:} 
\begin{align*}
Z_{\CH}(M,\Gamma;&j_1,...,j_n)\\&=N^{g-K-1}\left<\CH(H_-,m,\e,\Phi) \cup \Phi(\G);\W,j_1,...,j_n \right>\\
&=N^{n_1-n_0-n_1-n_2+\frac{n_1+n_2+1}{2}}Z_\WRT(\d(M^{(2)} \times I),\Gamma \times \{0\};j_1,...,j_n)\\
&=N^{-\frac{\chi(M) +n_3}{2}}Z_\WRT(\d(M^{(2)} \times I),\Gamma \times \{0\};j_1,...,j_n);\\
\end{align*}
here $n_i$ is the number of $i$-handles of $M$,
note that $M$ has a unique $0$-handle, by assumption. Now, $\d (
M^{(2) }\times I)=\d(M \times I) \# (S^1 \times S^2)^{ \# n_3}$, since
$M$ has a non-empty boundary.  The manifold $\d(M \times I)$ is exactly
the double $D(M)$ of $M$, in other words $M \cup_{\d M} \overline{M}$. 
Thence, from remark \ref{WRTp}: 
\begin{Theorem}  
Let $M$ be a connected compact {oriented} piecewise linear 3-manifold. Let also $\G$ be a trivalent framed  graph embedded in $M$, coloured with the spins $j_1,...,j_n \in \{0,1/2,...,(r-2)/2\}$. Choose a handle decomposition of $M$ such that $\G$ is contained in the union of the 0- and 1-handles of $M$. The Chain-Mail Expectation Value:
$$   {Z_{\CH}(M,\Gamma;j_1,...,j_n)=N^{g-K-1}\left<\CH(H_-,m,\e,\Phi) \cup \Phi(\G);\W,j_1,...,j_n \right>}$$
does not depend on the handle decomposition of $M$ containing $\G$ in the union of the 0- and 1-handles, and moreover it  is an invariant of the pair $(M,\Gamma)$, up to orientation preserving diffeomorphism. In fact:
$$   {Z_{\CH}(M,\Gamma;j_1,...,j_n)= N^{-\frac{\chi(M)} {2}}Z_\WRT(D(M),\Gamma;j_1,...,j_n ),}$$   {where $\G$ is embedded  in $M \subset D(M)$ in the obvious way.}
\end{Theorem}
Note that we have already proved this theorem for the case of closed manifolds.

Considering the case when $\G$ is the empty graph, it follows that the
Chain-Mail Invariant $Z_\CH(M)$ is also well defined for a manifold with
boundary.    {See also \cite{BP}.} This gives    {a definition} of the Turaev-Viro Invariant $Z_\TV(M)$ for manifolds with    {boundary;}    {cf. \cite{KMS}.}

\subsubsection{Calculation of the Invariants Using Triangulations-I}\label{TViro}

Let $M$ be  3-dimensional closed {connected} oriented piecewise linear manifold. Consider a piecewise linear triangulation of $M$. We can consider a  handle decomposition of $M$ where each $i$-simplex of $M$ generates a $(3-i)$-handle of $M$. See for example \cite{R1,R2}. Applying the Chain-Mail picture to this handle decomposition, yields the following combinatorial  picture for the  calculation of the Chain-Mail Invariant $Z_{\CH}(M)$, which, in this form is called  the Turaev-Viro Invariant $Z_\TV$.

\begin{figure}
\centerline{\relabelbox 
\epsfysize 10cm
\epsfbox{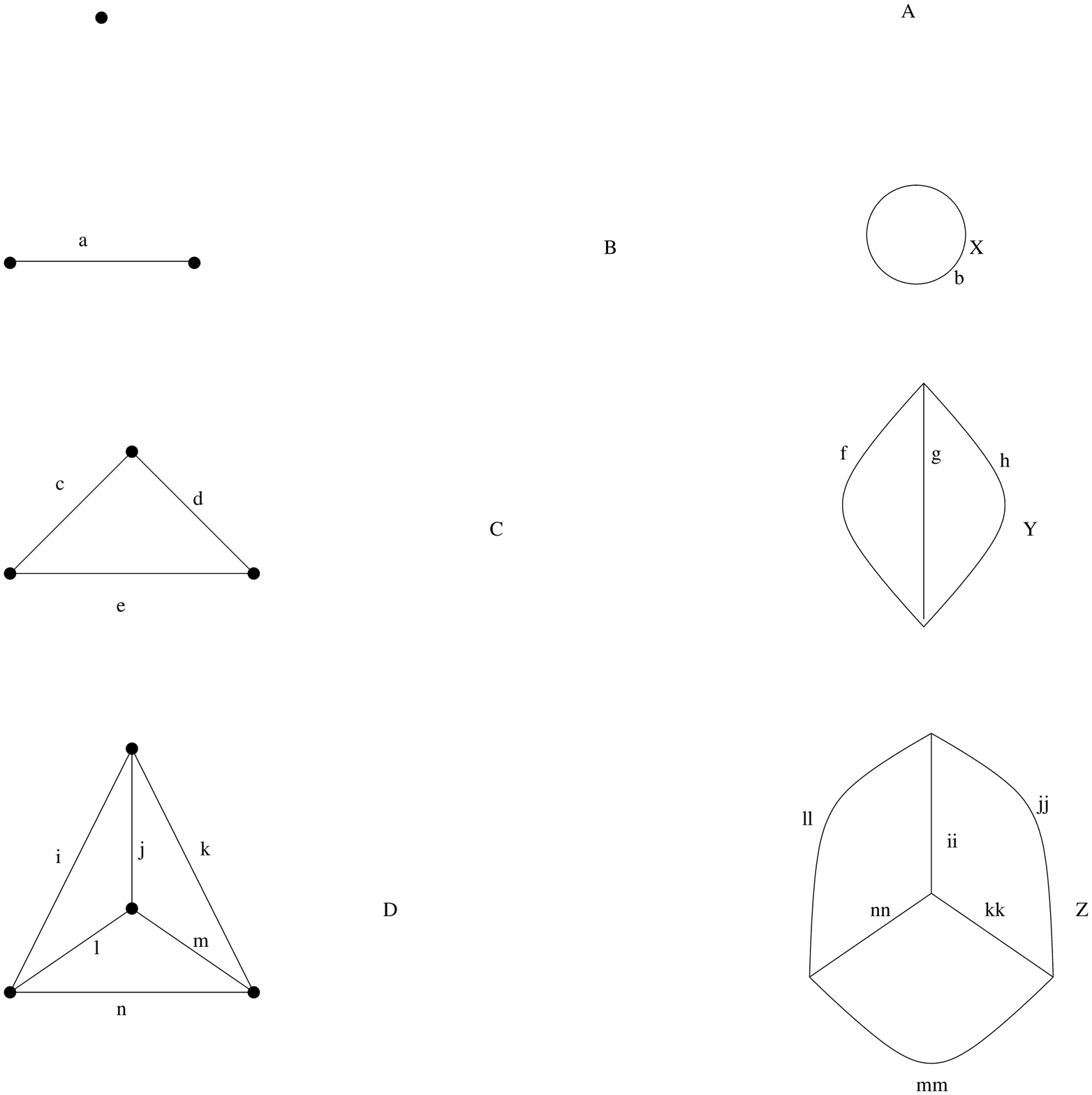}
\relabel{A}{$N^{-1}$}
\relabel{a}{$\s{a}$}
\relabel{B}{${\dim_q(a)=}\big  \langle $}
\relabel{b}{$\s{a}$}
\relabel{c}{$\s{c}$}
\relabel{d}{$\s{d}$}
\relabel{e}{$\s{e}$}
\relabel{f}{$\s{c}$}
\relabel{g}{$\s{d}$}
\relabel{h}{$\s{e}$}
\relabel{i}{$\s{i}$}
\relabel{j}{$\s{j}$}
\relabel{k}{$\s{k}$}
\relabel{l}{$\s{l}$}
\relabel{m}{$\s{m}$}
\relabel{n}{$\s{n}$}
\relabel{C}{${\theta(c,d,e)^{-1}=}\Big \langle$}
\relabel{ii}{$\s{i}$}
\relabel{jj}{$\s{j}$}
\relabel{kk}{$\s{k}$}
\relabel{ll}{$\s{l}$}
\relabel{mm}{$\s{m}$}
\relabel{nn}{$\s{n}$}
\relabel{D}{$\tau(i,j,k,l,m,n)= \Big \langle$}
\relabel{X}{$  \quad   \big \rangle$}
\relabel{Y}{$ \quad  \Big \rangle^{-1}$}
\relabel{Z}{$ \quad  \Big \rangle$}
\endrelabelbox }
\caption{\label{Weight} Weights associated with coloured simplices. All spin  networks are given the blackboard framing. }
\end{figure}

A colouring of $M$ is an assignment of a spin    {$j \in \{0,1/2,...,(r-2)/2\}$} to each edge of $M$. Each colouring of a simplex $s$ gives rise to a weight $W(s)$, in the way shown in figure \ref{Weight}.

\begin{figure}
\centerline{\relabelbox 
\epsfysize 4cm
\epsfbox{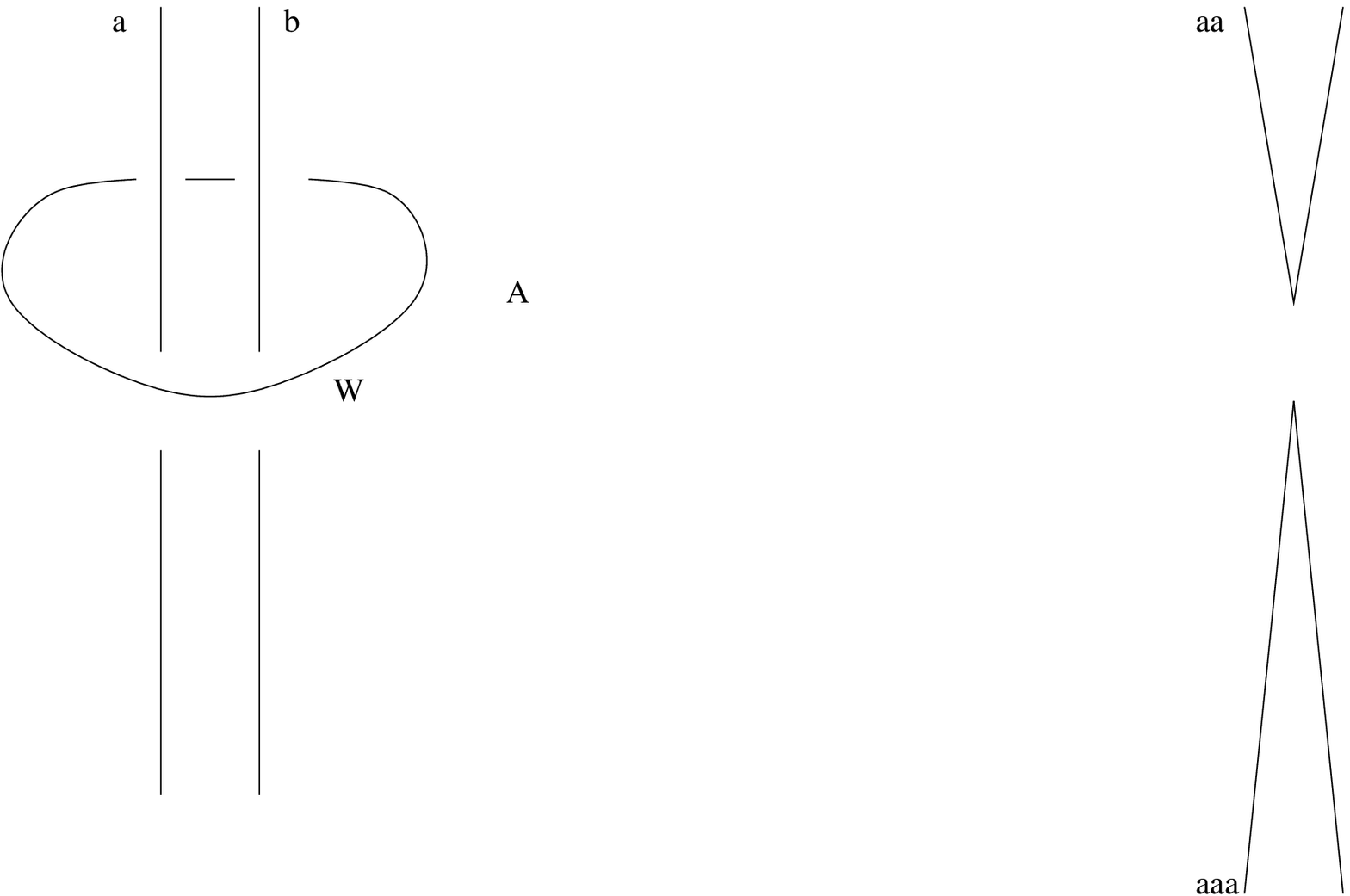}
\relabel{W}{$\s{\W}$}
\relabel{a}{$\s{a}$}
\relabel{aa}{$\s{a}$}
\relabel{aaa}{$\s{a}$}
\relabel{b}{$\s{b}$}
\relabel{A}{$=\frac{N}{\dim_q(a)}\delta(a,b)$}
\endrelabelbox } \caption{\label{Lic2} Lickorish Encircling Lemma for the case of two strands. All networks are given the blackboard framing. }
\end{figure}

\begin{figure}
\centerline{\relabelbox 
\epsfysize 4cm
\epsfbox{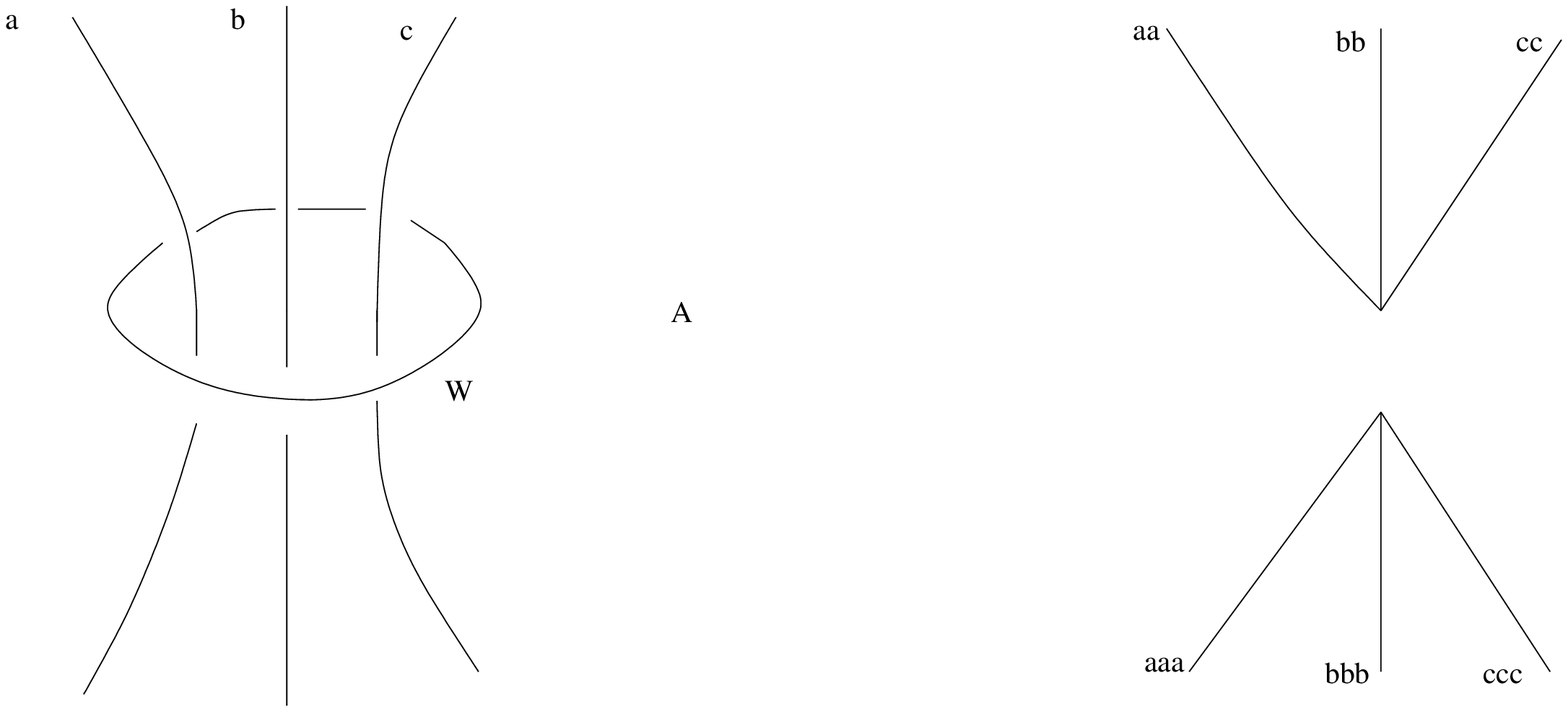}
\relabel{W}{$\s{\W}$}
\relabel{a}{$\s{a}$}
\relabel{b}{$\s{b}$}
\relabel{c}{$\s{c}$}
\relabel{aa}{$\s{a}$}
\relabel{bb}{$\s{b}$}
\relabel{cc}{$\s{c}$}
\relabel{aaa}{$\s{a}$}
\relabel{bbb}{$\s{b}$}
\relabel{ccc}{$\s{c}$}
\relabel{A}{$=N\theta(a,b,c)^{-1}$}
\endrelabelbox }\caption{\label{lic} Lickorish Encircling Lemma for the case of three strands. All networks are given the blackboard framing. }
\end{figure}
\begin{figure}
\centerline{\relabelbox 
\epsfysize 6cm
\epsfbox{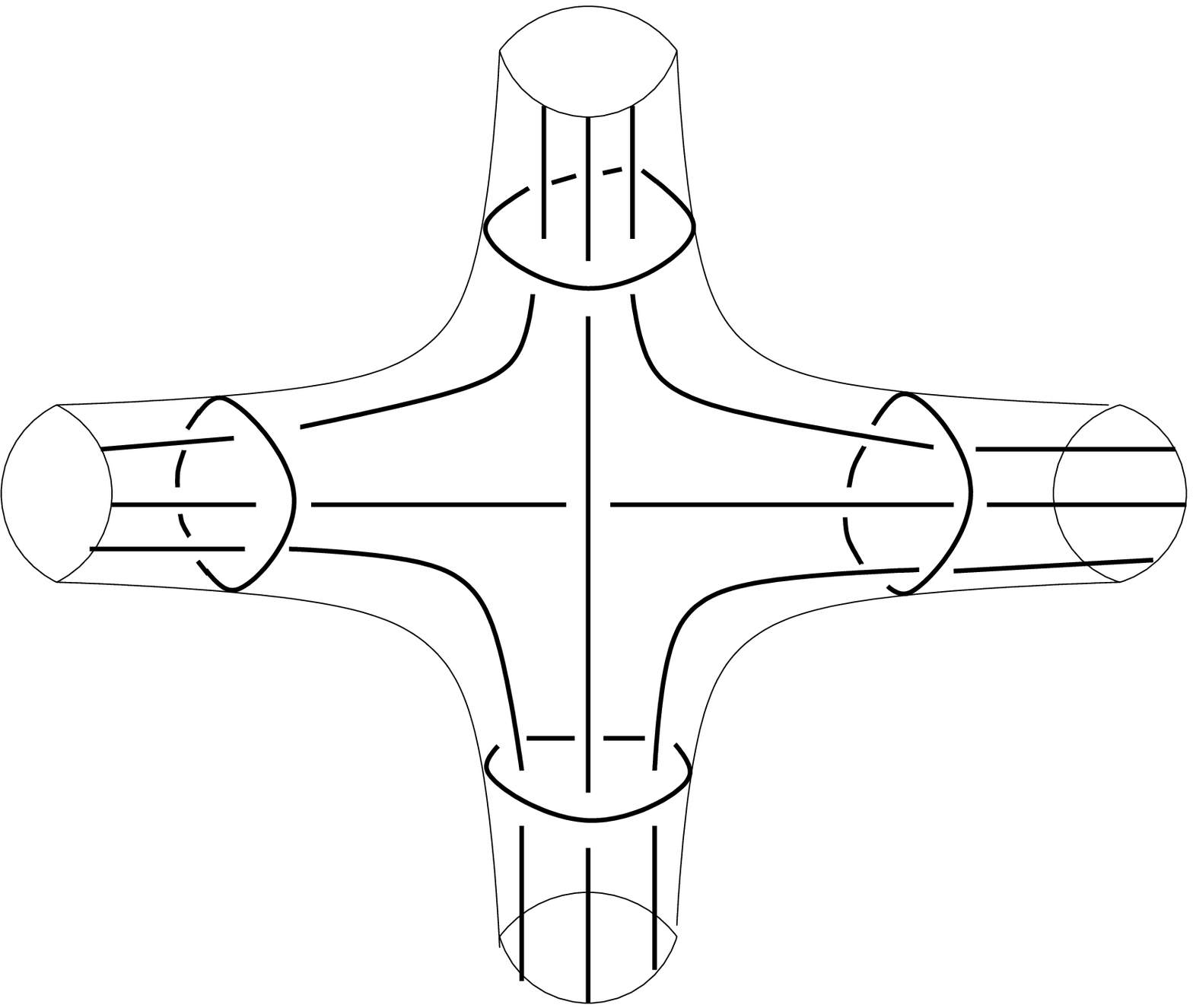}
\endrelabelbox }\caption{\label{Tetrahedron} Local configuration of the Chain-Mail Link at the vicinity of a tetrahedron. }
\end{figure}
\begin{figure}
\centerline{\relabelbox 
\epsfysize 4cm
\epsfbox{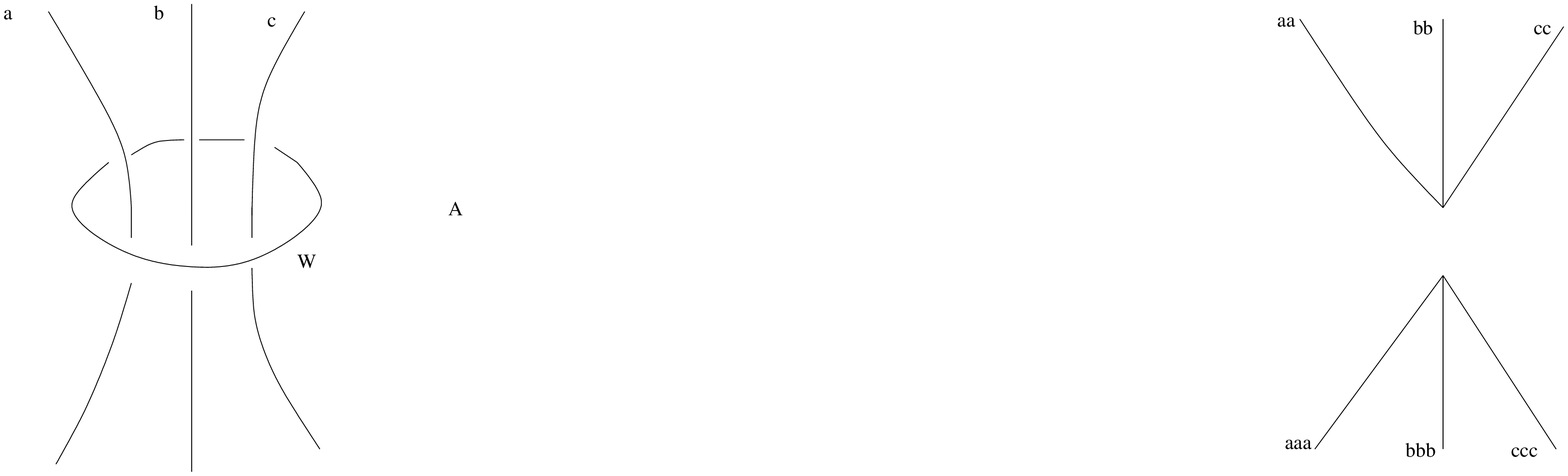}
\relabel{W}{$\s{\W}$}
\relabel{a}{$\s{\W}$}
\relabel{b}{$\s{\W}$}
\relabel{c}{$\s{\W}$}
\relabel{aa}{$\s{a}$}
\relabel{bb}{$\s{b}$}
\relabel{cc}{$\s{c}$}
\relabel{aaa}{$\s{a}$}
\relabel{bbb}{$\s{b}$}
\relabel{ccc}{$\s{c}$}
\relabel{A}{$=\displaystyle{N\sum_{a,b,c} \dim_q(a) \dim_q(b) \dim_q(c)\theta(a,b,c)^{-1}}$}
\endrelabelbox }\caption{\label{arg}Applying Lickorish Encircling Lemma to the 0-framed link determined by a face of the triangulation of $M$. }
\end{figure}

Using the identity shown in figure \ref{lic} and figure \ref{Tetrahedron}, it follows that:
\begin{align*}
Z_{\CH}(M)&=\sum_{\textrm{colourings}} \quad \prod_{\textrm{simplices } s} W(s)\\ &=Z_{\TV}(M).
\end{align*}
Note that we apply Lickorish Encircling Lemma to the 0-framed unknot defined from each face of the triangulation of $M$; see figure \ref{arg}.  
The last expression for $Z_{\CH}$ is the usual definition of the Turaev-Viro Invariant.  For a complete proof of the fact  that $Z_{\TV}=Z_\CH$,  see \cite{R1,R2}. Note  that we need to use the fact that the  Euler characteristic of any closed oriented 3-manifold is zero to obtain this. 
It is important to note that this argument only works if $M$ has an
empty boundary. If not, some extra terms will appear on the boundary of
$M$; see \ref{triang}.

\begin{figure}
\centerline{\relabelbox 
\epsfysize 4cm
\epsfbox{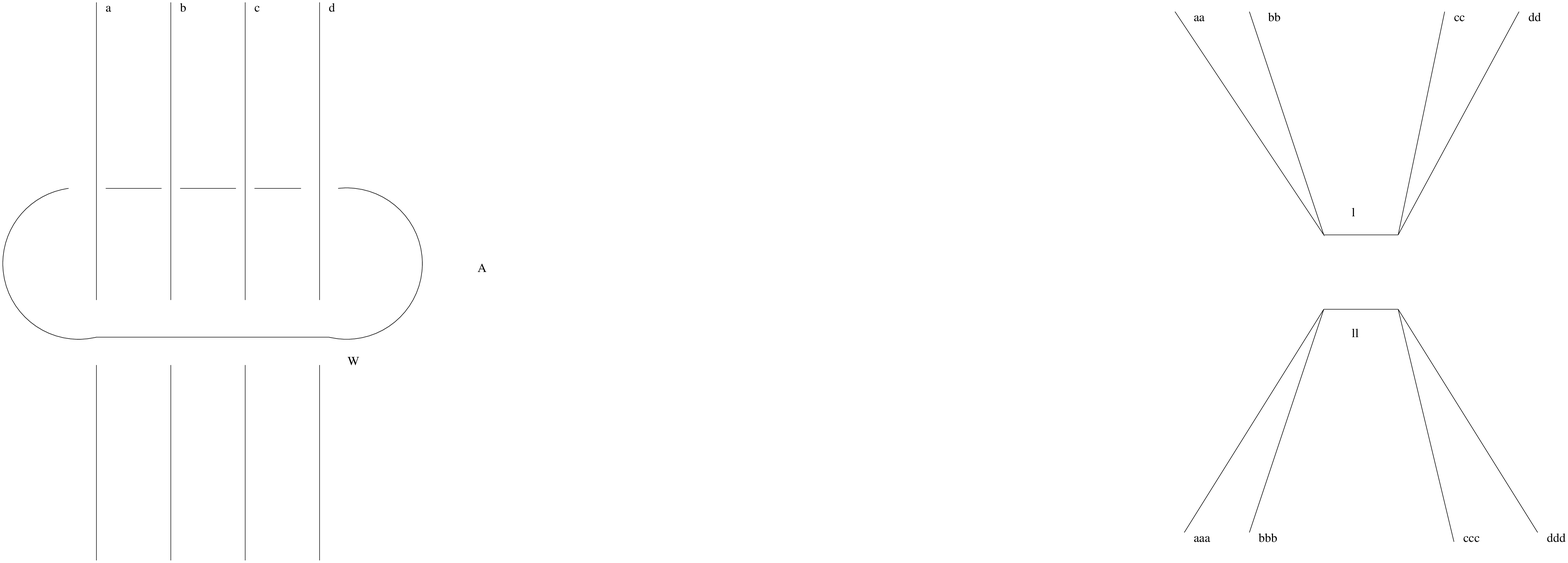}
\relabel{W}{$\s{\W}$}
\relabel{a}{$\s{a}$}
\relabel{b}{$\s{b}$}
\relabel{c}{$\s{c}$}
\relabel{d}{$\s{d}$}
\relabel{aa}{$\s{a}$}
\relabel{bb}{$\s{b}$}
\relabel{cc}{$\s{c}$}
\relabel{dd}{$\s{d}$}
\relabel{aaa}{$\s{a}$}
\relabel{bbb}{$\s{b}$}
\relabel{ccc}{$\s{c}$}
\relabel{ddd}{$\s{d}$}
\relabel{l}{$\s{l}$}
\relabel{ll}{$\s{l}$}
\relabel{A}{$=   {N\displaystyle{\sum_{l} \dim_q (l)} \theta(a,b,l)\theta(c,d,l)}$}
\endrelabelbox }\caption{\label{lic4} General  Lickorish Encircling Lemma for the case of four strands. All networks are given the blackboard framing. }
\end{figure}

Suppose  that $\G$ is a coloured trivalent framed graph  embedded in
the closed triangulated 3-manifold $M$. Suppose also that $\Gamma$ is contained in
the handlebody $H_-$  made from  the    {0- and {1-handles}} of $M$, where $M$ is given the handle decomposition associated with a triangulation of $M$.  Then the previous argument can be applied to obtain a combinatorial expression for the Chain-Mail Expectation Value  $Z_{\CH}(M,\G;j_1,...,j_n)$.  Anytime a strand of the graph $\G$ passes though a face of the triangulation of $M$, then we must apply a general Lickorish Encircling Lemma. See figure \ref{lic4} for the case of $4$ strands.

    {In this way one can express $Z_\CH(M,\G;j_1,..,j_n)$ as a state sum    {of} a product of $nj$ symbols.  This state sum appeared originally in \cite{M1}, and as described in the introduction section, it was obtained by a quantum group regularisation of the group integral (\ref{pid}). In fact, the Lickorish Encircling Lemma can be understood as a property of the integral over the quantum group of a product of quantum group representations, see \cite{O}.}

Later (see \ref{triang}) we will present a different  combinatorial expression for  $Z_\CH(M,\G;j_1,..,j_n)$ if $M$ is a piecewise-linear manifold with a piecewise-linear triangulation.

\subsection{Turaev's Description of the Invariants}
\subsubsection{Combinatorial Expression of the Observables in the Case of Thick Surfaces}\label{Thick}

Let $\S$ be an oriented surface. Let also  $\G$ be a  coloured trivalent framed graph embedded in $\S \times I$. Let us find a combinatorial expression for {the Chain-Mail Expectation Value} $Z_{\CH}(\S \times I,\G;j_1,...,j_n)$. Consider a handle decomposition of $\G$ constructed in the following way:

First of all, project the graph $\Gamma$ onto a graph $G$ in $\S\cong \S
\times \{1\}$,  so that $G$ is a regular projection of $\G$. Note that
we need to keep track of under- and over-crossings of $\G$.  By
definition, the point $(x,a)$ where $x \in \S$ and $a \in [0,1]$ will
be below  $(x,b)$ if $a<b$, here $b \in [0,1]$.  The graph  $G$ is in
general not 3-valent, since it will be 4-valent at the crossing
points of $\G$.

Second,  add  some extra edges and vertices to $G$, and obtain a graph $G^*$, such that $\S \setminus  G^*$ is a disjoint union of disks, and, moreover, no edge of $G^*$ closes up to a loop $S^1$; see \cite[proof of proposition 3.15]{R2}.

Therefore, there exists a handle decomposition of $\S \times I$ for which the union of the 0- and 1-handles of it is a thickening $n(G^*)$ of $G^*$. In particular, each vertex of $G^*$ induces a 0-handle of $\S \times I$ and each edge of $G^*$ induces a 1-handle of $\S \times I$.  There exists also a 2-handle for each connected component of $\S \setminus G^*$. Note that $\G$ is included in $n(G^*)$ in the obvious way. 

Each face $f$ of $\S \setminus G^*$ induces 2-handle of $\S \times I$, thus an element $l_f$ of the Chain-Mail Link. Let $l_F$ be the union of all these links, where $F$ is the set of faces of $\S \setminus G^*$. A face colouring $f \mapsto c_f$ is a map $F \to \{0,1/2,...,(r-2)/2\}$.  Each face colouring induces a colouring $c_F$ of $l_F$ by spins in $\{0,1/2,...,(r-2)/2\}$. 

Let $m$ be the meridian link of the 1-handles of $\S \times I$. Therefore, each edge of $G^*$ induces an element of $m$. Consider an orientation preserving map $\Phi\colon    \S\times I \to S^3$ which factors through an unknotted embedding of a handlebody of index 1. We have:
\begin{align*} 
Z_{\CH}&(\S\times {I}, \Gamma; j_1,...,j_n)=N^{ -\# \{ \textrm{vertices  of } G^*\}-\# \{ \textrm {faces  of }\S \setminus G^*\}}\\&\sum_{\textrm{face colourings of } \S \setminus G^*}  \left <\Phi (m) \cup \Phi (l_F)  \cup \Phi(\G);\W,c_F,j_1,...,j_n\right> \prod_{f \in F} \dim_q(c_f). 
\end{align*}
Now,  apply    {Lickorish Encircling Lemma} to each edge of $G^*$. This will    {yield} an additional factor of $N$    {for} each edge of $G^*$,    {in the previous formula}.      {Applying it firstly to  the edges of $G^*$ which are not in $G$  makes it clear} that, in order that a face colouring  of $\S \setminus G^*$  do not    {evaluate to} a trivial contribution  for $Z_{\CH}(\S\times    {I}, \Gamma; j_1,...,j_n)$, then for each  face of $\S \setminus G$, the colours  of  all faces of $\S \setminus G^*$ contained in it    must  coincide. This follows from figure \ref{Lic2}.    {Note that $G^*$ is obtained from $G$ by adding some extra edges, the thickening of each of them is  bearing  a 0-framed unknot coloured with the $\W$-element.}

\begin{figure}
\centerline{\relabelbox 
\epsfysize 12cm
\epsfbox{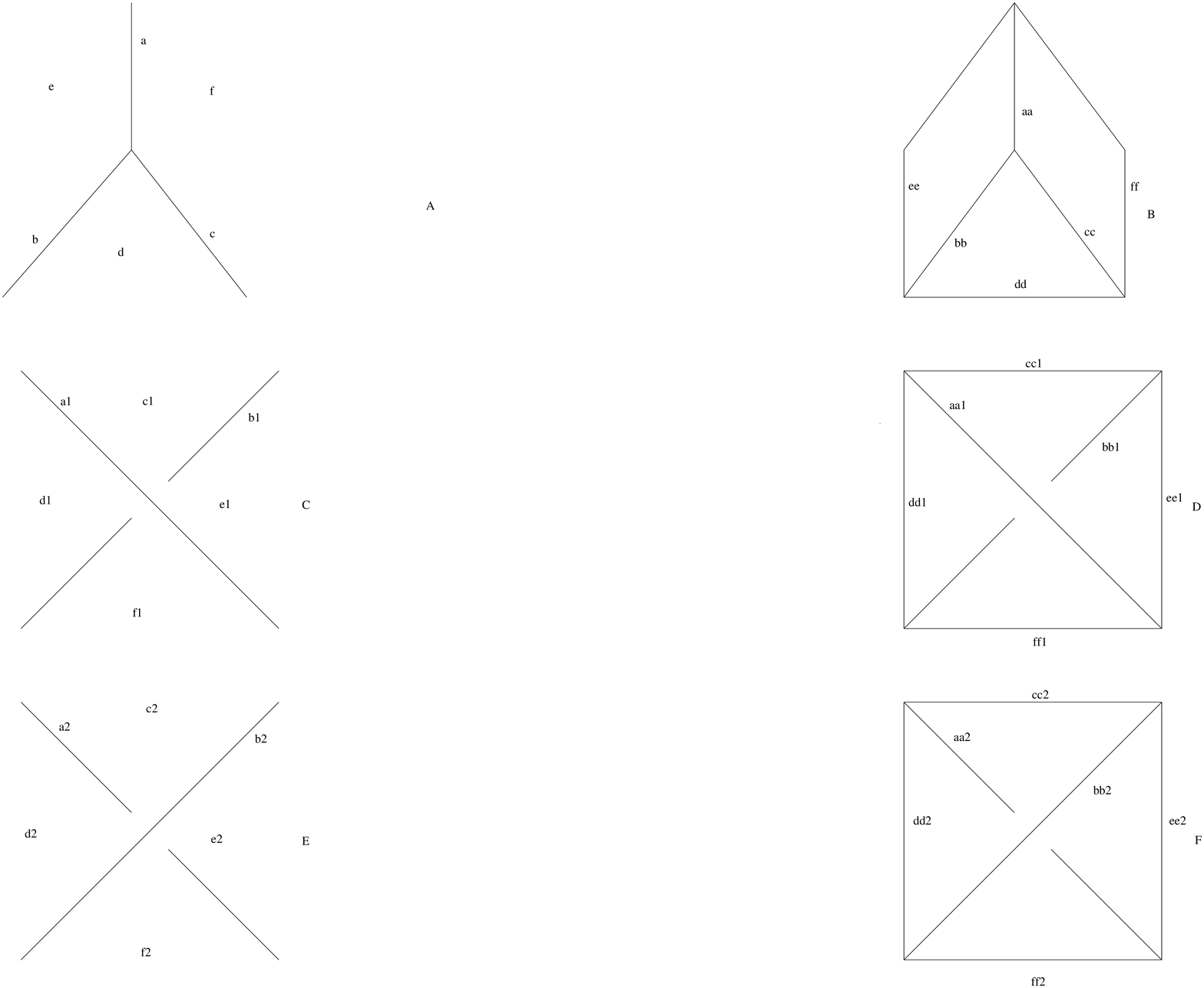}
\relabel{a}{$\s{a}$}
\relabel{b}{$\s{b}$}
\relabel{c}{$\s{c}$}
\relabel{d}{$\s{d}$}
\relabel{e}{$\s{e}$}
\relabel{f}{$\s{f}$}
\relabel{aa}{$\s{a}$}
\relabel{bb}{$\s{b}$}
\relabel{cc}{$\s{c}$}
\relabel{dd}{$\s{d}$}
\relabel{ee}{$\s{e}$}
\relabel{ff}{$\s{f}$}
\relabel{aa1}{$\s{a}$}
\relabel{bb1}{$\s{b}$}
\relabel{cc1}{$\s{c}$}
\relabel{dd1}{$\s{d}$}
\relabel{ee1}{$\s{e}$}
\relabel{ff1}{$\s{f}$}
\relabel{a2}{$\s{a}$}
\relabel{b2}{$\s{b}$}
\relabel{c2}{$\s{c}$}
\relabel{d2}{$\s{d}$}
\relabel{e2}{$\s{e}$}
\relabel{f2}{$\s{f}$}
\relabel{a1}{$\s{a}$}
\relabel{b1}{$\s{b}$}
\relabel{c1}{$\s{c}$}
\relabel{d1}{$\s{d}$}
\relabel{e1}{$\s{e}$}
\relabel{f1}{$\s{f}$}
\relabel{aa2}{$\s{a}$}
\relabel{bb2}{$\s{b}$}
\relabel{cc2}{$\s{c}$}
\relabel{dd2}{$\s{d}$}
\relabel{ee2}{$\s{e}$}
\relabel{ff2}{$\s{f}$}
\relabel{A}{${\left (\theta(e,a,f)\theta(f,c,d) \theta(e,b,d)\right)^{-\frac{1}{2}} \Big {\langle}}$}  
\relabel{C}{${\left (\theta(d,a,c)\theta(c,b,e) \theta(e,f,a)\theta(f,b,d)\right)^{-\frac{1}{2}} \Big {\langle}}$} 
\relabel{B}{$\Big{\rangle}$}
\relabel{D}{$\Big{\rangle}$}
\relabel{F}{$\Big{\rangle}$}
\relabel{E}{${\left (\theta(d,a,c)\theta(c,b,e) \theta(e,f,a)\theta(f,b,d)\right)^{-\frac{1}{2}} \Big {\langle}}$} 
\endrelabelbox }
\caption{\label{weight2} Weights associated with vertices of a graph projection on a surface $\S$. }
\end{figure}

Therefore we define a  colouring $c$ of $(\S,G)$ as being  a map from the set of connected
components of $\S \setminus G$ with spins $j\in \{0,1/2,...,(r-2)/2\}$. Note
that the edges of $G$  have colourings induced by the colourings of
the edges of $\Gamma$. Given a vertex $v$ of $G$, we assign a weight
$W(v,c)$ to it in the fashion shown in figure \ref{weight2}.    {Note that even though we need to choose a square root for the evaluation of each $\theta$-net, it is obvious from our construction that the final result will not depend on this choice.}

Similarly to the proof of the fact  that $Z_{\TV}=Z_{\CH}$ we obtain:

 \begin{multline*}
 Z_{\CH}(\S \times I, \Gamma;j_1,...,j_n)=N^{ -\# \{ \textrm{vertices  of } G^*\}+\# \{ \textrm{edges  of } G^*\}-\# \{ \textrm {faces  of }\S \setminus G^*\}}  \\ \sum_{\textrm{colourings } c  \textrm{ of } (\S,G)}  \quad\left ( \prod_{\textrm{vertices } v \textrm{ of } G}  W(v,c) \right) \left (\prod_{\textrm{faces } f \textrm{ of } \S \setminus  G} \left (\dim_q(c_f) \right)^{\chi(f)}\right),  
\end{multline*} 
 where  $c_f$ denotes the colouring of the face $f$, and $\chi(f)$ is the Euler characteristic of it.     {We refer to \cite[3.5]{R2} for details, in the case when $\S=S^2$.}

    {The reason for the factor $\left (\dim_q(c_f) \right)^{\chi(f)}$ is because after applying Lickorish Encircling Lemma in figure \ref{Lic2} to each edge of $G^*$ not in $G$, there is still  a contribution, which is, for each face $f$ of $\S \setminus G$, the factor $\dim_q(c_f)^{k_2-k_1+k_0}$, where $k_1$ is the number of    {edges} of $G^*$ which are not in $G$ and are included in the face $f$, $k_2$ is the number of faces of $\S \setminus G^*$ contained in $f$, and finally $k_0$ is the number of vertices of $G^*$ not in $G$ which are in the interior of $f$. Now $k_2-k_1+k_0=\chi(f)$. To see this note that $f$ has a natural handle decomposition with $k_2$ 0-handles,  $k_1$ 1-handles and $k_0$ 2-handles.}

   {Therefore if follows that:} 
$$Z_\CH(\S \times I, \Gamma;j_1,...,j_n)=N^{-\chi(\S)} Z_\T(\S \times I, \Gamma;j_1,...,j_n),$$
where
\begin{multline*}
Z_\T(\S \times I, \Gamma;j_1,...,j_n)=\\ \sum_{\textrm{colourings } c  \textrm{ of } (\S,G)}  \quad \left (\prod_{\textrm{vertices } v \textrm{ of } G}  W(v,c)\right) \left (\prod_{\textrm{faces } f \textrm{ of } \S \setminus  G} \left (\dim_q(c_f) \right)^{\chi(f)}\right).
\end{multline*}
The invariant  $Z_T$ of coloured graphs in thick surfaces $\S \times I$ was originally defined by V. Turaev; see  \cite{T3,T4}. Note that our normalisation is different.

\subsubsection{Invariants of Graphs Embedded in Index-1 Handlebodies}\label{invhand}
Let $H$ be a 3-dimensional oriented handlebody made from 0- and  1-handles. Consider the link $ m'$ made from the meridian curves  (the belt spheres) of  the 1-handles of $H$. 

Suppose that $\G$ is  a coloured framed  graph embedded  in $H$. Consider an orientation preserving embedding $\Phi\colon   H \to S^3$. 
Then, by definition: 
\begin{align*}
Z_\CH(H,\Gamma;j_1,...,j_n)&=N^{g-n_1-1} \left < \Phi(m') \cup \Phi(\Gamma); \W, j_1,...,j_n\right >\\&=N^{-n_0} \left < \Phi(m') \cup \Phi(\Gamma); \W, j_1,...,j_n\right >,
\end{align*}   
where $n_0$ and $n_1$ are  the number of 0- and  1-handles of the chosen
handle decomposition of $H$, and $g$ is the genus of the boundary of $H$; in particular $n_1=\# m'$. 

Another way to calculate $Z_\CH(H,\Gamma;j_1,...,j_n)$ is to consider the
combinatorial construction of the Chain-Mail Expectation Value for a graph embedded in $\d H \times I$ of \ref{Thick}. Let us explain this. First of all, choose an orientation of the surface of $\Gamma$ (like in \cite[5.a]{T3}). If we are provided with such an orientation, then we can shift $\G$ as close as we want to  $\d H$, by using the normal vector field to the surface of $\G$ determined from  the orientation (by construction, the final result will not depend on the orientation chosen).  

We resume the notation and discussion of \ref{Thick}.
Recall that we can define a handle decomposition of $\d H \times I$ for which
the union of the 0- and 1-handles of it is a thickening of the graph $G^*$. To
obtain a handle decomposition of $H$, we will need to add to $G^*$ the
projection of the meridians $m'$ of the 1-handles of $H$, which should stay
below $G^*$. This defines  a  graph $J$ living in in $\d H$.  If necessary,
add  some extra edges and vertices to $J$, obtaining a graph $J^*$ such that
the complement of $J^*$ in $\d H$ is a disjoint union of disks, and, moreover, no edge of $J^*$ closes up to a loop.

 Then, we attach a  2-handle along each circle of $m'$, and finally    {complete} the
 final result with 3-handles. This yields a handle decomposition of $H$, whose
 union of $0$- and 1-handles is a thickening $n(J^*)$ of $J^*$ in $H$.

Of
 course, there is also defined  a handle decomposition of $\d H\times I$, and
 the only difference is that the handle decomposition of $H$ has $\# m'$ extra
 $2$-handles, and also some 3-handles.  Therefore, the Chain-Mail Link
 associated with the constructed handle decomposition of $H$ has $\# m'$ more
 components than the Chain-Mail Link of the handle decomposition of $\d H \times
 I $.  Let $m$ be the meridian link of the handle decomposition of $\d H
 \times I$, and $\e$ the link defined from the  attaching regions of the
 2-handles of $\d H \times I$.

Consider an orientation preserving map $\Phi\colon    n(J^*) \to S^3$.    {We have:}
\begin{align*}
Z_\CH(H,\Gamma;j_1,...,j_n)&= N^{g-1-K-\# m'}\left <\Phi(m \cup m' \cup \e\cup \Gamma); \W,\W,\W,j_1,...,j_n  \right>\\
                          &= N^{-\# m'} Z_\CH(   {\d H \times I},\Gamma \cup m';j_1,...,j_n, \W).
\end{align*} 
Here $g$ is the genus of $\d( n(J^*))$ and $K$ is the number of $1$- and
$2$-handles of the handle decomposition of $\d H \times I$.     {In particular:}

$$ Z_{\CH}(H, \Gamma; j_1,...,j_n)=N^{-\chi(\d H) -\# m'} Z_\T(   {\d H \times I}, \Gamma \cup m'; j_1,...,j_n, \W),$$
by the discussion in \ref{Thick}.
\subsubsection{Calculation Using Triangulations-II}\label{triang}
Let $M$ be an oriented    {connected} closed piecewise linear 3-manifold. Let also $\G$ be a framed graph embedded in $M$. Suppose that $\G$ is coloured with the spins $j_1,...,j_n$.

We present now a fully combinatorial way to evaluate the Chain-Mail
Expectation Value $Z_{\CH}(M,\G;j_1,...,j_n)$, due to V. Turaev. This was in fact the original definition. See \cite{T2,T3,T4}.

Let $M'$ be $M$ minus a regular neighbourhood  $n(\G)$ of $\G$. This regular neighbourhood has a handle decomposition where each vertex of $\Gamma$ induces a $0$-handle and each edge of $\Gamma$ a 1-handle of $n(\Gamma)$. If necessary, we still need to add up some extra bivalent vertices of $\G$, so that no edge of $\G$ closes up to a loop.    {This yields a framed graph $\G'$}. Let $m'$ be the meridian link of this handle decomposition of $n(\Gamma)   {=n(\G')}$. 

Recall the construction of the Turaev-Viro Invariant in \ref{TViro}. Consider a triangulation of $M'$. This triangulation restricts to a triangulation of the boundary of the regular neighbourhood $n(\G)$ of $\G$. Consider the dual graph $G$ to this triangulation of $\d  (n(\G))$, pushed inside $n(\G)$, slightly, with framing parallel to the surface of $n(\G)$. Any colouring $c$ of the edges of $M'$ by spins induces a colouring $c_G$ of this graph  $G$ in $\d (n(\G))$.

   {Let $H=n(\Gamma)$}. Define:
\begin{align*}Z&_\TV (M,\Gamma;j_1,...,j_n)\\&=\sum_{\textrm{colourings  } c  \textrm{ of } M'} \quad Z_\CH(H,\G\cup G;j_1,...,j_n, c_G)\prod_{\textrm{simplices } s\textrm{ of } M'} W(s)\\&=\sum_{\textrm{colourings  } c  \textrm{ of } M'} \quad N^{   {-\chi(\d (H))} -\# m'}  Z_\T(   {\d H\times I},\G\cup G \cup m';j_1,...,j_n, c_G,\W)\\ &\quad\quad\quad\quad\quad\quad\quad\quad\quad\quad\quad\quad\quad\quad\quad \quad\quad\quad\quad\quad\quad\quad\quad\quad\prod_{\textrm{simplices } s\textrm{ of } M'} W(s).
\end{align*}
 This coincides with Turaev's definition    {in \cite{T2,T3,T4}}, apart from normalisation.
\begin{Theorem}
We have:

  $$Z_\TV (M,\Gamma;j_1,...,j_n)=N^{   {-\chi(M')}}Z_\CH (M,\Gamma;j_1,...,j_n).$$   

\end{Theorem} 

\begin{Proof}
   {As above, choose a handle decomposition of $n(\G)$ for which each vertex of $\Gamma'$
  induces a $0$-handle of $n(\Gamma)=n(\G')$ and each edge of $\Gamma'$ a 1-handle of it.} 
  This handle decomposition of $n(\G) \subset M$ can be completed to a handle
  decomposition of $M$ by using the triangulation of $M'$, in the usual
  picture where an $i$-simplex of $M'$ generates an $(3-i)$-handle of $M$.  As
  usual, we denote  the number of $i$-simplices of the triangulation of $M'$
  by $n_i$.

 Let $m$ be the meridian link of the 1-handles of $M$, minus the meridian link $m'$ of    {$n(\G')$}. Let also $\e$ be the    {framed} link obtained from the attaching regions of the 2-handles of $M$.  Consider the Chain-Mail expression for  $Z_\CH (M,\Gamma;j_1,...,j_n)$ using this handle decomposition, thus:
\begin{multline*}
Z_\CH(M,\G;j_1,...,j_n)\\=N^{n_2 -n_3+\#\{\textrm{edges of } \G' \}-\#\{\textrm{vertices of } \G' \}-\#\{\textrm{edges of } \G' \}-n_1-n_2}\\ \left <\Phi(m) \cup\Phi( m') \cup \Phi(\e) \cup \Phi(\G); \W,\W,\W,j_1,...,j_n \right>.
\end{multline*}
Here $\Phi$ is an orientation preserving embedding from the handlebody made from the 0- and 1-handles of $M$ into $S^3$.  By     {applying Lickorish Encircling Lemma}    {to the circles determined by the  triangles of $M'$,} in the same way as in the proof that $Z_{\CH}=Z_\TV$, this is  the same as: 

\begin{multline*}
Z_\CH(M,\G;j_1,...,j_n)=N^{-n_3-n_1-\#\{\textrm{vertices of } \G' \}+n_2+n_0}\\ \sum_{\textrm{colourings  } c  \textrm{ of } M'} \quad    {\left <\Phi(\G) \cup \Phi(G)\cup \Phi(m');j_1,...,j_n, c_G,\W \right >}\prod_{\textrm{simplices } s\textrm{ of } M'} W(s),
\end{multline*}
or

\begin{align*} Z_\CH&(M,\G;j_1,...,j_n)
\\&=N^{   {\chi(M')}} \sum_{\textrm{colourings  } c  \textrm{ of } M'}  Z_\CH(H,\G\cup G;j_1,...,j_n, c_G)\prod_{\textrm{simplices } s\textrm{ of } M'} W(s)
\\ &=N^{   {\chi(M')}}Z_{\TV}(M,\Gamma;j_1,...,j_n).
\end{align*}

\end{Proof}

\noindent   {Considering empty graphs, this result also  provides a new
  proof of the fact that the combinatorially defined  Turaev-Viro Invariant
  can  be extended to  3-manifolds with boundary;    {see also \cite{BP}}. Compare with \cite{KS,KMS,BD,T3,T4},  where similar invariants of graphs were considered.}

\section{Concluding remarks}

Since $Z_\CH(M,\Gamma)$ is based on the path integral  (\ref{pid}), it is straightforward to define a    {state sum expression for} $Z_\CH(M,\Gamma)$ for higher-dimensional manifolds. It would    {be} interesting to investigate    {whether there exists} a higher-dimensional analog  the relation (\ref{bfcs}).     {In the two dimensional case,}  the path integral (\ref{pid}) can    {also} be regularised by using  quantum groups,    {see}  \cite{M3}, and our results imply that the corresponding state sum is given by    { $Z_{\CH}(\Sigma \times I,\G)$.}    {These invariants will be relevant for three-dimensional Euclidean LQG as well as for topological string theory}. Note that in two dimensions the path integral (\ref{pid})
is convergent for    {surfaces of  non-zero genus} and empty graphs    {(see \cite{FK,Ba})}, which then    {raises the} possibility of defining the    {corresponding} spin network invariants without    {using  quantum groups.} Such invariants will be observables    {of two-dimensional Quantum Gravity} with zero cosmological constant.    {For} the case of three-dimensional Quantum Gravity with zero cosmological constant see \cite{FL}.

As shown in \cite{M2}, in order to construct more realistic quantum gravity states one needs to consider more general wavefunctions. The    {associated} spin-network coefficients will correspond to the chain-mail     {evaluations considered in this article, but} with additional loops    {carrying insertions}.  

The relation (\ref{bfcs}) should be helpful    {for understanding} the loop transforms of both the Kodama state and the flat connection state. According to the proposal of \cite{FS}, the loop transform coefficients of the Kodama state should be given by $Z_{\WRT}(M,\Gamma)_{i\cdot k}$, where $Z_{\WRT}(M,\Gamma)_k$ is the Witten-Reshetikhin-Turaev Invariant for the case when    { $q=e^{\frac{i \pi}{k+2}}$,} while the cosmological constant would be proportional to $1/k$. On the other hand, according to the proposal of \cite{M1},    {and the results in this article,} the loop transform coefficients of the flat-connection state in the Euclidean case are given by $Z_{\WRT}(M,\G)_k Z_{\WRT}({\overline{M}})_k$, where $k$ is taken to be sufficiently large in order to obtain a vanishingly small, i.e. zero, cosmological constant. Note that these two loop transforms are essentially the same in the Euclidean case. However, extending them to the Minkowski case and obtaining a complete understanding of the role of the cosmological constant, are still open problems.

\section*{Acknowledgements}
   {
J. Faria Martins was financed by  Funda\c{c}\~{a}o para a Ci\^{e}ncia e
Tecnologia (Portugal), post-doctoral grant number SFRH/BPD/17552/2004, part of
the research project POCTI/MAT/60352/2004 (``Quantum Topology''), also
financed by the FCT. A. Mikovi\'c was partially supported by the FCT grant POCTI/MAT/45306/2002.
We would like to  thank John W. Barrett for introducing us to some  of the techniques used in this article and
Vladimir Turaev for giving us the exact references for his work on invariants of coloured graphs.}

\end{document}